\documentclass[12pt]{article}
\usepackage{latexsym, amssymb, amsmath}

 \def\pd{\partial}  \def\a{\alpha} \def\b{\beta} \def\dl{\delta} \def\s{\sigma}    \def\lam{\lambda} \def\Lam{\Lambda}  \def\Gm{\Gamma} \def\om{\omega}  \def\sq{\sqrt} \def\fr{\frac} \def\half{\frac{1}{2}}
 
 \def\hg{{\hat g}} \def\bg{{\bar g}}  
\def\nb{\nabla}    
\def\bDelta{{\bar \Delta}}

\def\lap3{~| \!\!\! \partial^2} \def\dlap3{~| \!\!\! \partial^4} \def\invlap3{~| \!\!\! \partial^{-2}}
\def\lang{\langle} \def\rang{\rangle} 

  \def\pl{{\rm pl}}

\begin{document}

\begin{center}
{\large {\bf Revealing A Trans-Planckian World Solves The Cosmological Constant Problem}}
\end{center}

\begin{center}
{\sc Ken-ji Hamada}
\end{center}

\begin{center}
{\it Institute of Particle and Nuclear Studies, KEK, Tsukuba 305-0801, Japan  \\ and
Department of Particle and Nuclear Physics, The Graduate University for Advanced Studies (SOKENDAI), Tsukuba 305-0801, Japan}
\end{center}

\begin{abstract}
The Planck scale is usually believed to be an unpassable wall. Putting a cutoff there and thinking of it as a quantized spacetime entity shows that. However, this is exactly the cause of many problems in quantum gravity. The cosmological constant problem also comes down to the problem of how to describe a trans-Planckian world by a continuum theory that shall be renormalizable and background free. Here, we show that when quantizing gravity in a diffeomorphism-invariant method, in which background freedom arises asymptotically as an exact conformal symmetry, the zero-point energy vanishes identically. Thus, there is no problem with the cosmological constant, which is a physical constant, namely a renormalization group invariant. We also argue that new quanta based on the background freedom emerge and spacetime is discretized dynamically. Primordial fluctuations will originate from such quanta.
\end{abstract}

\section{Introduction}

In the late 1990s, observations of distant Type Ia supernovae led by Perlmutter,  Riess, and Schmidt \cite{rs, perlmutter} confirmed that the universe is currently undergoing accelerated expansion, the results of which can be explained by the cosmological constant. Until the cosmological constant was discovered, the question of its existence was a major issue, although  its very existence is not a problem in classical gravity because it does not conflict with diffeomorphism invariance at all. The problem is that, on proceeding to quantum field theory and calculating quantum corrections to the cosmological constant, a large value of the Planck mass to the fourth power is yielded, which is about 120 orders of magnitude different from the observed value \cite{weinberg89}.

The concrete calculation was first made by Zel'dovich \cite{zeldovich}. He introduced an ultraviolet (UV) cutoff in the Planck scale to evaluate the zero-point energy and derived a particularly large value. The reason for introducing the UV cutoff to the Planck scale is that Einstein gravity is believed to be able to describe spacetime correctly at least up to that region. If the calculated value has physical meaning, then the presence of the cutoff is also physical. In other words, it claims that a world discretized by the Planck length, $l_{\rm pl} = 1.616 \times 10^{-35}\,{\rm m}$, is an entity of quantum spacetime.

The view of spacetime discretized by the Planck length is often seen when considering quantization of gravity. A big problem with this idea is that diffeomorphism invariance is broken. Despite the fact that what predicts the existence of the cosmological term is diffeomorphism invariance, if the quantization method breaks the invariance, it is a wrong way for the end.

On the other hand, if we quantize gravity presuming that fields exist continuously even in the world beyond the Planck scale and diffeomorphism invariance is preserved there, we can show that there is no zero-point energy by a Schwinger-Dyson (SD) equation, which represents background-metric independence as follows:\footnote{
We consider the universe to be a four-dimensional spacetime without boundaries.} 
\begin{eqnarray}
    \int [dg] \fr{\dl}{\dl g_{\mu\nu}(x)} e^{iI} 
       =  i \fr{1}{2} \lang \sq{-g} \,T^{\mu\nu} (x) \rang  = 0 ,
       \label{gravitational SD equation}
\end{eqnarray}
where $I$ is an action consisting of the gravitational field $g_{\mu\nu}$ and other fields (omitted here), and $T^{\mu\nu}$ is the whole energy-momentum tensor. This identity indicates that the cosmological constant problem will be solved if the path integral for the gravitational field can be defined correctly in a diffeomorphism-invariant way.

This is a general conclusion, but does not hold for the method of introducing the cutoff as a finite value. Therefore, it is necessary to reveal how to define the path integral in the region beyond the Planck scale. It is also a matter of spacetime singularity, renormalizability, and unitarity. In this paper, we revisit these problems, which have been studied individually so far \cite{adm1}--\cite{book}, with quantum gravity and make clear that they are deeply related to the cosmological constant problem. We then argue that the cosmological constant is nothing but a physical constant, i.e., a true constant whose value does not change during the evolution of the universe.

\section{Into A Trans-Planckian World}

First of all, a point-like elementary particle is a concept that contradicts Einstein gravity. Still, if mass $m$ is smaller than the Planck mass $m_{\rm pl}$, it can be approximated as a particle because the Compton wavelength $\lam \sim 1/m$ is longer than the horizon size $h_g \sim m/m_\pl^2$ created by the mass itself. However, if $m > m_\pl$, then $\lam < h_g$, and its information is confined inside the horizon and lost. Hence, the ordinary particle picture is no longer valid in the world beyond the Planck scale.

Moreover, Einstein gravity is unstable because the Einstein-Hilbert (EH) action is unbounded below; thus the path integral is ill-defined and cannot handle strong gravity phases. Although we can apply a weak field approximation, it becomes unrenormalizable \cite{dn, tv74, veltman, weinberg79}. Besides that, we cannot eliminate singularities because the action is finite for them as is the case for physical objects.\footnote{
This is easy to understand by considering Euclidean theory Wick-rotated in a background introduced later. If an action $I$ is finite such as for solitons and instantons, then the path integral weight $e^{-I}$ is also finite, while, if $I$ positively diverges, the weight disappears, so that its existence probability vanishes.} 
This is tied to the fact that, since the Einstein equation does not contain the Riemann tensor,  it has insufficient ability to control spacetime curvature, unlike gauge field equations involving field strengths.

Since the late 1970s, attempts have been made to modify Einstein gravity by adding fourth-derivative actions in order to make it renormalizable \cite{stelle, tomboulis1, tomboulis2, ft}. The modified action also has the following significant properties: bounded below and no singularities. The former is due to the curvature being squared, which removes the instability. The latter is a property that holds if the Riemann tensor squared is added, in which case the action diverges positively for singularities as unphysical.\footnote{
The Riemann tensor squared is often removed from the action using the Euler combination, but this can only be done when it is clear that there is no singularity in spacetime, such as when considering perturbation theory around a regular spacetime. Here we are discussing general cases without such a premise.} 

Despite these good properties, it still does not work. First, the action cannot be uniquely determined only by these. More problematic is the quantization method. The attempt was made with a weak field expansion around vanishing Riemann tensor, i.e., with a normal particle picture, but what we are looking for is a theory describing a trans-Planckian world that would be in a strong gravity phase. In fact, applying this approximation causes the so-called ghost problem. Although there is an idea \cite{stelle, tomboulis1, tomboulis2, ft} based on Lee and Wick's work \cite{lw} to make ghosts unphysical by interaction effects, it has not led to an essential solution to the problem. Something is missing to overcome such physical and theoretical problems.

A hint to solve the ghost problem is in the early universe. Cosmologists support inflation theory that there was a period in which space expanded rapidly before the big bang in order to solve the horizon and flatness problems. If we believe in the typical inflation scenario proposed by Guth \cite{guth}, Sato \cite{sato}, and Starobinsky \cite{starobinsky}, then most of the universe that can be observed today was born from a range narrower than the Planck length. Inflation is usually described by a conformally flat de Sitter spacetime; hence it suggests that quantum spacetime will be described by a perturbation theory around a vanishing Weyl tensor.

Therefore, when defining such a perturbation, the conformal factor, which is the most important factor determining distance, is extracted in an exponential form so as to be positive, and is treated non-perturbatively like
\begin{eqnarray}
    g_{\mu\nu} = e^{2\phi} \bg_{\mu\nu} ,
       \label{metric expansion about conformal flat}
\end{eqnarray}
where the scalar-like mode $\phi$ is called the conformal-factor field. The remaining modes that represent deviation from the conformally flat spacetime are expanded by introducing a dimensionless coupling constant $t$ as $\bg_{\mu\nu} = \bigl( \hg e^{th} \bigr)_{\mu\nu} = \hg_{\mu\lam} \, (\dl^\lam_{~\nu} + t h^\lam_{~\nu}  + t^2 h^\lam_{~\s} h^\s_{~\nu}/2  + \cdots )$, where $h^\mu_{~\nu}$ is a traceless tensor field and $\hg_{\mu\nu}$ is a background metric with signature $(-,+,+,+)$.

Since the field strength of the traceless tensor field is the Weyl tensor, the perturbation theory is formulated as follows \cite{hs}--\cite{hm17}:
\begin{eqnarray}
   I = \int d^4 x \sq{-g} \, \biggl[  - \fr{1}{t^2} C^2_{\mu\nu\lam\s} - b G_4 
       + \fr{1}{\hbar} \biggl( \fr{1}{16\pi G} R -\Lam + {\cal L}_{\rm M} \biggr) \biggr] ,
    \label{background free quantum gravity action}
\end{eqnarray}
where $C_{\mu\nu\lam\s}$ is the Weyl tensor and $G_4=R^2_{\mu\nu\lam\s}-4R^2_{\mu\nu} + R^2$ is the Euler density. The first two are conformally invariant, and $b$ is not an independent coupling constant because the Euler term has no kinetic term. The third is the EH action, and the fourth is the cosmological term that has to be positive from boundedness at low energy. These depend on $\phi$ exponentially. ${\cal L}_{\rm M}$ is a matter term that is conformal at high energy. For simplicity, bare and renormalized quantities are not distinguished. Note that since the gravitational field is dimensionless unlike other fields, the fourth-derivative gravitational actions become exactly dimensionless; thus there is no $\hbar$ in front of them, which implies that they contribute only to quantum dynamics.

The point of quantization is to rewrite the theory into a quantum field theory defined on a background, and then show independence of how to choose the background. The partition function is then expressed as \cite{riegert}--\cite{hm17}
\begin{eqnarray}
    \int [dg] \, e^{iI(g)} = \int [d\phi dh]_\hg \,e^{i S(\phi,\bg)+i I(g)} ,
        \label{quantum gravity in any background}
\end{eqnarray}
where the $\phi$-dependent $S$ is the Wess-Zumino (WZ) action \cite{wz} that is necessary for rewriting the invariant measure $[dg]$ to an ordinary measure $[d\phi dh]_\hg$ defined using the background metric $\hg_{\mu\nu}$ \cite{polyakov, kpz, david, dk}.\footnote{
Diffeomorphism invariance demands that $S$ must satisfy a certain condition described in the following. Let us consider a simultaneous change of $\hg_{\mu\nu} \to e^{2\om} \hg_{\mu\nu}$ and $\phi \to \phi - \om$. The partition function (\ref{quantum gravity in any background}) is then invariant by definition because $g_{\mu\nu} \,$(\ref{metric expansion about conformal flat}) is invariant under this change. On the other hand, the measure on the right-hand side changes to $[d\phi]_{e^{2\om} \hg}[dh]_{e^{2\om} \hg} \, e^{S(\phi-\om, e^{2\om} \bg)} = [d\phi]_\hg [dh]_\hg \, e^{iS(\om,\bg)} \, e^{S(\phi-\om, e^{2\om} \bg)}$, where $[d\phi]_\hg$ is invariant under the shift of $\phi$ for any background because its integration region is $(-\infty,\infty)$. In order for this to return to the original form, $S$ has to satisfy $S(\om,\bg)+S(\phi-\om, e^{2\om} \bg)=S(\phi,\bg)$. This is called the WZ consistency condition, and thus $S$ is called the WZ action. This condition was originally shown in curved spacetime. Quantizing gravity, the above property of $[d\phi]_\hg$ ultimately results in the background-metric independence for the conformal change of $\hg_{\mu\nu}$.} 
A conformal variation of $S$ yields conformal anomaly \cite{cd}--\cite{hamada20b}. Historically, the term anomaly has been used, but physically it is never anomalous, which is necessary to preserve the diffeomorphism invariance. The SD equation (\ref{gravitational SD equation}) is then defined for each, so that the background-metric independence holds exactly for $\phi$ while perturbatively for $h^\mu_{~\nu}$.

The WZ actions are responsible for fourth-derivative dynamics of $\phi$, which basically arise linked with a scale introduced upon quantization. The most important is, however, a scale-invariant kinetic term remaining even in the UV limit of $t \to 0$, called the Riegert action \cite{riegert}: $- (b_1/16\pi^2) \int d^4 x \sq{-\hg} \, ( 2\phi \bDelta_4 \phi  + \bar{E}_4 \, \phi )$, where $E_4= G_4 -2\nb^2 R/3$ and $\sq{-g} \Delta_4$ is a conformally invariant fourth-order differential, defined by $\Delta_4 = \nb^4 + 2R^{\mu\nu} \nb_\mu \nb_\nu - 2R \nb^2/3 + \nb^\mu R \nb_\mu/3$. The quantity with the bar is the one defined by $\bg_{\mu\nu}$. The lowest of the coefficient $b_1$ is given by a positive number \cite{ft, amm92, hs}. At higher orders of $t$, WZ multipoint interactions such as $\phi^{n+1} (2 \bDelta_4 \phi + \bar{E}_4)$, $\phi^n \bar{C}_{\mu\nu\lam\s}^2$, and also $\phi^n {\bar F}_{\mu\nu}^2$ for a gauge field $F_{\mu\nu}$ with $n \geq1 $ arise \cite{hamada02, hamada14re, hamada20b}.

The Riegert action, together with the kinetic term of the Weyl action, describes a special conformal field theory (CFT) \cite{am, amm92, amm97, hh, hamada12a, hamada12b}. Here, it is somewhat confusing that, even though the Riegert action was originally derived as a conformal anomaly, it works to restore an exact conformal invariance when performing the path integral over $\phi$. This can be understood from the viewpoint of diffeomorphism invariance below.

The difference from normal CFT is that conformal invariance appears as part of diffeomorphism invariance $\dl_\xi g_{\mu\nu} = g_{\mu\lam} \nb_\nu \xi^\lam + g_{\nu\lam} \nb_\mu \xi^\lam$, namely a gauge symmetry or Becchi-Rouet-Stora-Tyutin (BRST) symmetry, in the UV limit \cite{hh, hamada12a, hamada12b}. That is, all theories with different backgrounds connected to each other by conformal transformations are gauge equivalent. This is an algebraic representation of the background-metric independence, called ``BRST conformal invariance", which was first studied in 2D quantum gravity \cite{polyakov, kpz, dk, david}. Therefore, employing the Minkowski background is justified, so that standard methods in field theory can be applied.

The transformation law for each mode is expressed as
\begin{eqnarray}
   \dl_\zeta \phi &=& \zeta^\lam \pd_\lam \phi + \fr{1}{4} \pd_\lam \zeta^\lam ,
             \nonumber \\
   \dl_\zeta h_{\mu\nu} &=& \zeta^\lam \pd_\lam h_{\mu\nu} 
             + \half h_{\mu\lam} \Bigl( \pd_\nu \zeta^\lam - \pd^\lam \zeta_\nu \Bigr)
             + \half h_{\nu\lam} \Bigl( \pd_\mu \zeta^\lam - \pd^\lam \zeta_\mu \Bigr) ,
         \label{conformal transformations of gravitational fields}
\end{eqnarray}
where $\zeta^\lam$ is the part satisfying conformal Killing equations in the gauge parameter $\xi^\lam$, while the rest is gauge-fixed, for instance, by employing the radiation gauge. The presence of the shift term in the first expression shows that this has a diffeomorphism origin. Note that when $t \neq 0$, corrections by $t$ appear in the second and diffeomorphism deviates from the conformal transformation.

Hence, unlike ordinary perturbation theories, there are no free particles, or gravitons, asymptotically; thus the scattering matrix, which is a physical quantity premised on the existence of them, is not defined.

Physical quantities must be invariant under the conformal transformation (\ref{conformal transformations of gravitational fields}). Despite being the lowest in $t$, this gauge transformation depends on the fields, unlike the lowest of normal gauge transformations; thus the story about unitarity changes. Actually, modes making up the field mix with each other under the transformation. Ghost modes serve as necessary elements to form the closed BRST conformal algebra, but we can show that they all become gauge variant, namely unphysical. Furthermore, we can show that there are an infinite number of physical fields or states, given by real composite scalars \cite{amm97, hh, hamada12a, hamada12b}. Matter fields are also dressed by gravity so as to be real scalars. The reality of physical fields will be maintained even in correlation functions due to the positivity of the Riegert and Weyl actions.

The appearance of the BRST conformal invariance in the high-energy limit is called ``asymptotic background freedom". This property is also consistent with the fact that scale-invariant scalar fluctuations dominate in the early universe \cite{wmap13, planck}.

\section{What Diffeomorphism Invariance Shows}

Gravity is usually thought to occur where matter is present. To emphasize this, the Einstein equation is expressed by arranging the matter energy-momentum tensor $T^{\rm M}_{\mu\nu}$ as a source term. This convention, however, hides the essence of diffeomorphism invariance. The Einstein equation is the equation of motion for the gravitational field derived based on the variational principle. This means that the whole energy-momentum tensor vanishes as $T_{\mu\nu}=0$; in particular, the Hamiltonian disappears.

Going to quantum gravity, the equation of motion is expressed by the SD equation (\ref{gravitational SD equation}). The Hamiltonian and momentum constraints \cite{adm1, adm2}, and also the Wheeler-DeWitt equation \cite{dewitt1}--\cite{wheeler2} are essentially the same as this equation, though the terms are mostly used for Einstein gravity. The constraints have to form a closed algebra. The BRST conformal algebra is the first example of such a diffeomorphism algebra shown to be closed at the quantum level.

The SD equation is an expression of the background-metric independence. Changing the background metric is equivalent to changing the fields that are integral variables while preserving the full metric tensor (\ref{metric expansion about conformal flat}). That is to say that the equation of motion can be expressed as an equation in which variation of the effective action with respect to the background metric vanishes (see the appendix for the effective action). That of the quantum gravity has the following structure \cite{hy, hhy1, hhy2, hamada20a}: 
\begin{eqnarray}
     T^{(4)}_{\mu\nu}  + \fr{m_{\rm pl}^2}{8\pi} \biggl( - R_{\mu\nu} + \half g_{\mu\nu} R \biggr)
                        - \Lam_{\rm cos} \, g_{\mu\nu} + T^{\rm M}_{\mu\nu} = 0,
       \label{full energy momentum tensor}
\end{eqnarray}
where the second is the Einstein tensor term and $\Lam_{\rm cos}$ is the cosmological constant. The fourth-derivative term $T^{(4)}_{\mu\nu}$ involves contributions from not only the conformally invariant actions but also various quantum corrections, including the WZ actions for conformal anomalies.

As in the case of quantum chromodynamics (QCD), the effective action can be expressed in terms of a running coupling constant \cite{hamada20b, hy, hhy1, hhy2, hamada20a}:
\begin{eqnarray}
     {\bar t}^2(Q) = \bigl[ \b_0 \log (Q^2/\Lam^2_{\rm QG}) \bigr]^{-1} .
     \label{running coupling}
\end{eqnarray}
The coefficient $\b_0$ is from the beta function that is negative as in $\mu d t/d\mu = -\b_0 t^3 + o(t^5)$ \cite{ft, amm92, hs}, where $\mu$ is an arbitrary mass scale introduced upon quantization. At this time, the momentum squared $Q^2$ depends on the conformal-factor field $\phi$ so that the effective action is diffeomorphism invariant, and this dependence is exactly the WZ action for conformal anomaly. The new scale $\Lam_{\rm QG}\, (= \mu e^{-1/2\b_0 t^2})$ is a dynamical scale of quantum gravity, which appears as a renormalization group (RG) invariant \cite{collins}.

Now, let us describe what the vanishing of the whole energy-momentum tensor means in physics. Two important things are addressed below:

\paragraph{I.} 
One is that, since the Hamiltonian vanishes exactly, there is no absolute time globally defined in the whole system, unlike ordinary quantum field theory. This is also true in classical gravity. Time progresses differently depending on the strength of the gravitational field. In quantum spacetime where gravity is fully fluctuating, the concept of time itself disappears. Measurements of physical time and distance defined by the metric tensor (\ref{metric expansion about conformal flat}) are virtually impossible. The asymptotic background freedom refers to the realization of such spacetime in the UV limit.

The concept of conservation will be expressed as RG invariance. The energy-momentum tensor is conserved in this sense. Entropy of the universe, given by the effective action because whole energy vanishes, is also conserved.

Eventually, time is expressed as a dynamical change in gravity derived by solving the equation of motion. Monotonic increase of the conformal factor, called the scale factor in cosmology, represents the time flow of the entire universe. Inflation will be the first, and then the Friedmann solution brings time.

It should be noted here that, even though we are thinking of a solution in which the Hamiltonian vanishes, there is a dynamical solution. It comes from the fact that the EH action is not bounded below. If this action were positive-definite, similar to those of matter fields, then there would only be a trivial vacuum solution and no time would occur.

In quantum gravity, we have to solve the equation of motion (\ref{full energy momentum tensor}). Note that due to the presence of the $T^{(4)}_{\mu\nu}$ term, it makes sense even if the $T^{\rm M}_{\mu\nu}$ term disappears. This implies that it is possible to describe a world where there is no matter and only fluctuations of gravity exist. Further, this equation can describe a transition process from such a world to the world with matter, which will occur at the dynamical scale $\Lam_{\rm QG}$. These will be covered later.

\paragraph{II.}
Another important fact is that the equation of motion (\ref{full energy momentum tensor}), or Eq. (\ref{gravitational SD equation}), shows that there is no zero-point energy.

Before understanding this, it should be pointed out that the energy-momentum tensor is a normal product, a composite field that behaves as a finite operator in correlation functions. In ordinary quantum field theory, it can generally be shown as follows. Consider a partition function $Z = \int [d f] \, e^{iI(f,g)}$ for a field $f$ in curved spacetime. Since the partition function is finite, variation of it with respect to the metric tensor is obviously finite. Thus, we can find that the energy-momentum tensor containing quantum corrections is a finite operator as $\dl Z/\dl g_{\mu\nu}(x) = i \lang \sq{-g} \, T^{\mu\nu}(x) \rang/2 = {\rm finite}$. This is a consequence of diffeomorphism invariance.

Adler, Collins, and Duncan \cite{acd} have shown that $T^{\mu\nu}$ is actually expressed by a normal product, as part of the study of conformal anomalies using dimensional regularization that preserves diffeomorphism invariance manifestly \cite{tv72}, in which case information on the WZ action $S$ is contained between 4 and $D \,(<4)$ dimensions.\footnote{
When dimensional regularization is employed, the SD equation (\ref{gravitational SD equation}) holds as it is, including contributions from the measure, without rewriting as in Eq. (\ref{quantum gravity in any background}) \cite{hamada14re}.} 
Following this work, Brown and Collins \cite{bc} and Hathrell \cite{hathrell1, hathrell2} have derived special RG equations from the fact that the energy-momentum tensor is not only a normal product but also an RG invariant, and have shown, by solving the equations, that the Euler-type conformal anomaly is given exactly by the Riegert form $E_4$ \cite{hamada14a}. The quantum gravity action in dimensional regularization has been determined based on this result.

Further quantizing gravity, we find that the energy-momentum tensor vanishes identically as shown in Eq. (\ref{gravitational SD equation}). Therefore, we can conclude that there is no zero-point energy, so that the cause of the cosmological constant problem disappears.

In general, applying normal ordering to the Hamiltonian in field theories after canonical quantization may be a task to restore diffeomorphism invariance. This is because, unlike the path integral method, the canonical quantization method treats the time component specially, so it cannot be said that diffeomorphism invariance is manifestly preserved. In fact, it has been shown that the normal ordering is necessary to obtain the closed BRST conformal algebra \cite{amm97, hh, hamada12a, hamada12b}.

Then what is the cosmological constant $\Lam_{\rm cos}$ that appears in the equation of motion (\ref{full energy momentum tensor})? It is nothing but a physical constant. To be precise, it is an RG invariant satisfying $d\Lam_{\rm cos}/d\mu =0$, composed of renormalized quantities $\Lam$, $G$, $t$ in the action and also $\mu$ \cite{hm16, hm17}. Another example of such a physical constant is the dynamical scale $\Lam_{\rm QG}$ in Eq. (\ref{running coupling}); the Planck mass $m_{\rm pl}$ is another. These three constants must be determined from observations.

\section{Further Claims}

The most important factor is diffeomorphism invariance, which is the guiding principle even in quantum theory. The root cause of the cosmological constant problem is that this invariance is broken by the UV cutoff. In the first place, whether or not the problem arises depends on the method of regularization. Actually, there is no guarantee that the cutoff dependence will be in the form of the cosmological term. If we employ dimensional regularization, UV divergences corresponding to the power of the cutoff do not appear.

The disappearance of the zero-point energy affects problems with the origin of fluctuations in the early universe. In Einstein gravity, the source of the various types of matter that constitutes the present universe also has to be some kind of matter and, in many inflation models, its role is played by an unknown scalar field called the inflaton. The origin of fluctuations has then been explained as the zero-point energy of this field. However, diffeomorphism invariance denies the existence of such a zero-point energy; instead, the energy of the conformal gravity dynamics will serve as the source of everything, as seen below.

As inferred from QCD, the new dynamical scale $\Lam_{\rm QG}$ is considered to be a scale that separates quantum spacetime from the current classical spacetime. A big difference from QCD is that, even if the running coupling constant disappears, a kind of strong-coupling phase arises in which spacetime fluctuates so much that the background freedom is realized, i.e., not a phase with free particles like gravitons. Inversely, if the running coupling constant increases at the $\Lam_{\rm QG}$ scale, the fourth-derivative conformal gravity dynamics disappears, just as gauge field dynamics disappear at the QCD scale. At this time, since the whole energy-momentum tensor is preserved as zero as in Eq. (\ref{full energy momentum tensor}), the gravitational energy of the disappeared $T^{(4)}_{\mu\nu}$ is transferred to $T^{\rm M}_{\mu\nu}$. In this way, a spacetime phase transition will occur with the creation of matter.

We can then construct a Starobinsky-type inflation model characterized by the two scales $m_{\rm pl}$ and $\Lam_{\rm QG}$, which terminates with the spacetime phase transition as the big bang \cite{hy}. Scale-invariant gravitational fluctuations have been shown to decrease gradually as the running coupling constant increases during the inflation period, and the reduced ones are inherited to the Friedmann universe as its initial condition \cite{hhy1, hhy2}. This may lead to a solution to the Hubble tension \cite{hubble-tension} that appears to request a revision of the initial condition.

The magnitude of $\Lam_{\rm QG}$ has been estimated from the quantum gravity inflation scenario to be $10^{17}\,$GeV, which is two orders of magnitude smaller than the Planck mass. The correlation length of the quantum gravity is then $\xi_\Lam=1/\Lam_{\rm QG}$, about 100 times the Planck length. That is to say that, if we increase the energy to see a shorter distance, Einstein gravity breaks down before we reach the Planck scale, and spacetime enters the new phase. Therefore, we can solve many problems in Einstein gravity. The ordinary point-particle picture is valid only below $\Lam_{\rm QG}$, described by a derivative expansion around Einstein gravity, where $m_{\rm pl} > \Lam_{\rm QG}$ works as a unitarity condition to avoid reaching ghost poles.

The continuity demands a new perspective on spacetime quanta. The correlation length  $\xi_\Lam$ defines just such a quantum as a measurable minimum length. This is because, inside $\xi_\Lam$, quantum gravity is activated and background-free spacetime is realized, so that we are virtually unable to measure physical distance. Thus, the asymptotic background-freedom suggests that spacetime is dynamically discretized by $\xi_\Lam$, despite the theory being defined by a continuum theory. Localized excited states of quantum gravity with mass beyond the Planck mass are given by $\xi_\Lam$ in diameter and have a Schwarzschild tail outside \cite{hamada20a}.

\appendix
\section{More about Effective Action}

The beta function $\b_t =\mu dt/d\mu = -\b_0 \, t^3 + o(t^5)$ is negative and the 1-loop coefficient is given by $\b_0 = [ ( N_{\rm X} + 6 N_{\rm F}+ 12 N_{\rm A} )/240 + 197/60 ]/(4\pi)^2$, where $\mu$ is an arbitrary mass scale introduced upon quantization and $N_{\rm X}$, $N_{\rm F}$, and $N_{\rm A}$ are the number of scalar fields, Dirac fermions, and gauge fields belonging to the matter term ${\cal L}_{\rm M}$, respectively \cite{ft, amm92, hs, cd, ddi, duff1, duff2}. The corresponding Weyl sector effective action is, in momentum space, given by \cite{hamada02, hamada20b, hhy1, hhy2, hamada20a}
\begin{eqnarray}
    \Gm_{\rm W} &=& - \biggl[ \fr{1}{t^2} - 2 \b_0 \phi 
                                 + \b_0 \log \biggl( \fr{q^2}{\mu^2} \biggr) 
                                        \biggr] \sq{-\bg} \, {\bar C}^2_{\mu\nu\lam\s} 
                          \nonumber \\
                      &=& - \fr{1}{\bar{t}^2(Q)} \sq{-g} \, C^2_{\mu\nu\lam\s} ,
             \label{weyl effective action}
\end{eqnarray}
where $q^2$ is the momentum squared defined by the metric tensor ${\bar g}_{\mu\nu}$ excluding the conformal factor. The first part of the first line is the tree action, the second is exactly the WZ action, and the third is a nonlocal loop correction, which put together in the running coupling constant (\ref{running coupling}), where $Q^2=q^2/e^{2\phi}\,$. The dynamical energy scale is then expressed as $\Lam_{\rm QG}=\mu \, e^{-1/2\b_0 t^2}$. This is a renormalization group invariant satisfying $d \Lam_{\rm QG}/d\mu =0$. The same is true including higher order corrections, though expressions of the running coupling constant and the dynamical scale become more complicated \cite{hamada20b}.

The coefficient in front of the Riegert action is expanded as $b_1=b_c (1- a_1 t^2 + o(t^4))$, where the lowest value is given by $b_c = ( N_{\rm X} + 11 N_{\rm F}+ 62 N_{\rm A} )/360 + 769/180$ \cite{amm92, hs, cd, ddi, duff1, duff2}. The Riegert sector effective action will be described in a form so that $t^2$ is replaced with $\bar{t}^2(Q)$ as in the Weyl sector. This replacement indicates that the corresponding WZ multipoint interactions appear at higher orders \cite{hamada02, hamada20b, hhy1, hhy2, hamada20a}.

The physical cosmological constant satisfying $d\Lam_{\rm cos}/d\mu =0$, which is defined by the effective action $\Gm_\Lam = - \Lam_{\rm cos}\sq{-g}$, is given at the 1-loop level as follows \cite{hm17}:
\begin{eqnarray*}
    \Lam_{\rm cos} &=& \Lambda +  ( 7 - 2 \log 4\pi ) \frac{\Lambda}{b_c}
          - \left( \frac{\Lambda}{b_c} - \frac{9\pi^{2}M^{4}}{2b_c^2} \right)
            \! \log \! \left( \frac{64\pi^{2}}{\mu^{4}} \fr{\Lam}{b_c} \right)
                \nonumber \\
      &&  -  \fr{9\pi^2}{2} \! \left( \frac{25}{3} - 4 \log 4\pi   \right)  \! \frac{M^{4}}{b_c^2}
                  \nonumber \\
      &&  - 6 \pi \frac{M^{2}}{b_c}  \sqrt{\frac{\Lambda}{b_c} - \frac{9\pi^{2}M^{4}}{4b_c^2}} 
              \arccos  \! \left( \frac{3\pi M^{2}}{2\sqrt{b_c\Lambda}} \right) 
               \nonumber \\ 
      &&  + \frac{5}{128} \alpha_{t}^{2} M^{4} \! \left( \log \frac{\pi^2 \alpha_{t}^{2}M^{4}}{\mu^{4}} - \frac{21}{5} \right) ,
\end{eqnarray*}
where $\a_t=t^2/4\pi$. $M \!=\! 1/\sq{8\pi G}$ and $\Lam$ are renormalized quantities of the Planck mass and the cosmological constant in the action, respectively. To derive this expression, we take an approximation of large $b_c$, namely large number of matter fields, so that the ratios $\Lam/b_c$ and $M^4/b_c^2$ are comparable, as are $\a_t/4\pi$ and $1/b_c$.


\end{document}